\documentclass[letter]{aa} 

\usepackage[]{natbib}
\usepackage{graphicx}
\usepackage{txfonts}
\usepackage[lofdepth,lotdepth]{subfig}

 \usepackage{multirow}

\begin{document}

   \title{Origin of the ionized wind in MWC 349A}

   \author{A. B\'aez-Rubio
          \inst{1}
	  \and J. Mart\'in-Pintado\inst{1}
          \and C. Thum\inst{3}
	  \and P. Planesas\inst{2}
	  \and J. Torres-Redondo\inst{1}
          }

   \institute{Centro de Astrobiolog\'ia (CSIC-INTA),
              Ctra de Torrej\'on a Ajalvir, km 4, 28850 Torrej\'on de Ardoz, Madrid, Spain\\
              \email{baezra@cab.inta-csic.es, jmartin@cab.inta-csic.es}
         \and
		Observatorio Astron\'omico Nacional (IGN), Alfonso XII 3, E-28014 Madrid, Spain \\ \email{p.planesas@oan.es} 
	\and 
		Instituto de Radio Astronom\'ia Milim\'etrica (IRAM), Avenida Divina Pastora, 7, N\'ucleo Central, E 18012 Granada, Spain\\  \email{thum@iram.es}
             }

   \date{Received June 01, 2012; accepted June 01, 2012}

\abstract{The UC-HII region of MWC 349A is the prototype of an ionized wind driven by a massive star surrounded by a disk. Recent high angular resolution observations of the millimeter recombination lines have shown that the disk rotates with a Keplerian law in its outer parts. However, the kinematics of innermost regions in the UC-HII region of \mbox{MWC 349A} is still unknown, in particular the radius where the wind is launched from the disk. We performed hydrogen recombination line observations with the Heterodyne Instrument for the Far Infrared (HIFI) onboard the Herschel Space Observatory to study the kinematics of its innermost regions by studying their spectral features. In addition to the two laser peaks, we report the first detection of two new components that are blueshifted with respect to the laser peaks for all the recombination lines with principal quantum number $n\le21$. These new spectral features originate from the region where the wind is ejected from the disk. We used our 3D non-LTE radiative transfer model for recombination lines (MORELI) to show that these features are consistent with the wind being ejected at a radius of $\sim24$ AU from the star, which supports magnetohydrodynamic wind models.}


   \titlerunning{Tracing the ejection of ionized material in the innermost regions of MWC 349A using submm and FIR lasers}
   \authorrunning{B\'aez-Rubio et al.}
   \maketitle

\section{Introduction}

Circumstellar disks around massive stars play a key role in the formation of these stars through accretion, as predicted by the model of monolithic collapse \citep{McKee2003} and shown by observations (see e.g. \citealt{Cesaroni2007}). These dense disks are steadily photoionized by UV radiation from the stars, which leads to ionized outflows \citep{Jaffe1999}. However, the processes involved in the launching of the ionized wind are poorly known. 

Pure hydrodynamics models based on thermal pressure \citep{Hollenbach1994,Font2004} were discarded because of the region where the ionized winds are launched \citep{MartinPintado2011}. Alternatively, magnetohydrodynamics models explain the launching by magnetocentrifugal acceleration through magnetic field lines, whether in the innermost region of the disk threaded by the magnetosphere, as predicted by the X-wind models \citep{Shu1994a,Shu1994b,Shu2000,Ostriker1995}, or in an extensive range of radii, as predicted by the disk wind models \mbox{(e.g. \citealt{Blandford1982}).} However, it is still under debate which of these models better explains the observed steady winds. 

MWC 349A is the best example of a massive star with a neutral disk \citep{Danchi2001} powering an ionized outflow that expands at nearly constant velocity \citep{Olnon1975}. It is particularly interesting because it is the only UC-HII region, along with MonR2-IRS2 \citep{JimenezSerra2013}, with strong double-peaked hydrogen recombination lines (RLs) at wavelengths shorter than 3 mm (i.e. principal quantum numbers \mbox{$n\le39$,} \citealt{MartinPintado1989a}) arising from laser emission from a dense Keplerian-rotating ionized disk \citep{Planesas1992,Weintroub2008,MartinPintado2011}. 

A 3D radiative transfer modelling (MORELI code, \citealt{BaezRubio2013}) of the radio-continuum and the RL emission under non-LTE conditions indicated that the ionized wind might be launched from the ionized rotating disk, and its acceleration up to its terminal velocity, \mbox{60 km s$^{-1}$,} occurs very close to the disk. However, because of continuum opacity effects, revealing the kinematics of the innermost regions of the ionized disk and wind requires observations of RLs at higher frequencies. For this purpose, we have used the Heterodyne Instrument for the Far Infrared (HIFI) on the Herschel Space Observatory\footnote{Herschel is an ESA space observatory with science instruments provided by European-led Principal Investigator consortia and with important participation from NASA.} \citep{Pilbratt2010} to observe the whole set of RLs accessible with this instrument, that is, Hn$\alpha$ RLs with $15 \leq n\leq23$. Thanks to the high signal-to-noise ratio that is obtained in our observations, we reveal, for the first time, new kinematic features arising from the innermost regions of the disk from which the wind is ejected.

\section{Observations and data reduction}

We used the HIFI instrument \citep{deGraauw2010} onboard Herschel to acquire high spectral resolution profiles from all the RLs between the H23$\alpha$ and the H15$\alpha$ toward \mbox{MWC 349A} ($\alpha_{\mathrm{J2000}}=20^{\mathrm{h}}32^{\mathrm{m}}45^{\mathrm{s}}.54$, $\delta_{\mathrm{J2000}}=40\degr39'36.55''$). The spectra were taken by using simultaneously the Wide Band Spectrometer (WBS) and the High Resolution Spectrometer (HRS) in both horizontal and vertical polarization. The spectra from both polarizations were averaged to improve the signal-to-noise ratio. Since MWC 349A is a point-like source in the HIFI beam, we used the dual beam switch observing mode with fast chopping to obtain the best possible baselines for our broad RLs. We only used the slow chopper speed for the two most stable receivers, that is, for bands 1a and 1b to observe the H23$\alpha$ and the H22$\alpha$ RLs. Table \ref{observations} summarizes the observations. The data were reduced with the Herschel Interactive Processing Environment version 8.1.0 \citep{Ott2010}. The spectra were smoothed to a velocity resolution of \mbox{0.5 km s$^{-1}$,} except for the spectra of H15$\alpha$-H17$\alpha$, which were smoothed to 1 km s$^{-1}$ to improve the signal-to-noise ratio. The results found from both spectrometers are consistent, and we only show the profiles measured with the WBS. 

\begin{table}
\caption{HIFI observations toward MWC 349A}
\label{observations}
\centering
\begin{tabular}{ccccccc}  \hline
\vspace{-9.5pt}\\
Line & $\nu$ [GHz] & Band & Date &Observation ID \vspace{1pt} \\ \hline \hline 
\vspace{-7.5pt}\\
H23$\alpha$&507.18& 1a &2011 Nov 2& 1342231781\\ 
H22$\alpha$&577.90& 1b &2011 May 20&1342221419\\  
H21$\alpha$&662.40& 2a &2011 Nov 3&1342231802 \\ 
H20$\alpha$&764.23& 2b &2011 May 22&1342246007\\ 
H19$\alpha$&888.05& 3b &2011 May 20&1342221448\\ 
H18$\alpha$&1040.13& 4a&2011 May 20&1342221425\\
H17$\alpha$&1229.03& 5a&2011 Nov 3& 1342231784 \\ 
H16$\alpha$&1466.61& 6a&2011 Nov 3& 1342231800 \\ 
H15$\alpha$&1769.61& 7a&2011 May 20&1342221421\\ \hline 
\end{tabular}
\end{table}

\section{Results and discussion}
\label{results}
\subsection{Two new blueshifted kinematic features}

\begin{figure}
\centering
\includegraphics[width=8.1cm]{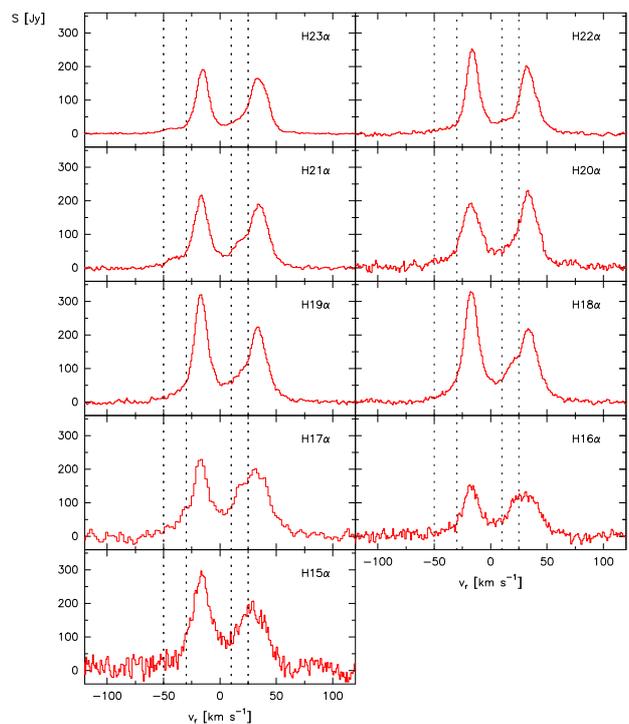}
\caption{Recombination line profiles detected with HIFI, with the continuum emission subtracted. The vertical dashed lines show the limits of the radial velocity ranges where the two blueshifted components are observed for Hn$\alpha$ with $n\le21$.}
\label{figure:RLs}
\end{figure}

\begin{figure}
\centering
\includegraphics[width=8.1cm]{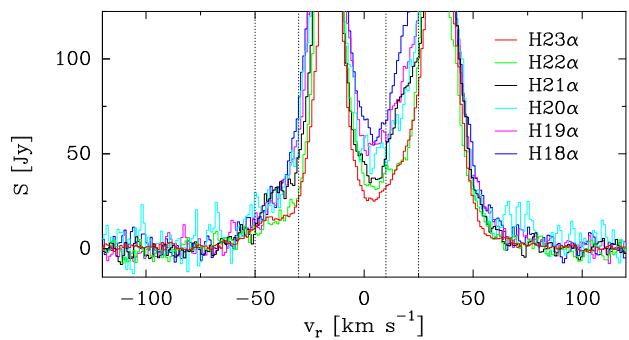}
\caption{Measured line profiles of Hn$\alpha$ RLs with $18 \leq n\leq23$. The vertical dashed lines indicate the limits of the radial velocity ranges of the two new kinematic components that are blueshifted with respect to the laser peaks. This figure clearly shows two new blueshifted kinematic components observed in the RLs with $n\leq21$.}
\label{H22alpha_and_H21alpha}
\end{figure}

The observed RL profiles are shown in Fig. \ref{figure:RLs}. They show the typical spectral features found in this source from lower frequency observations, that is, double-peaked profiles. In addition to those two well-studied laser peaks, we report two new kinematic features blueshifted by \mbox{$\sim$ 20 km s$^{-1}$} with respect to the two laser peaks (specifically, at radial velocities between -50 to \mbox{-30 km s$^{-1}$} and from 10 to 25 \mbox{km s$^{-1}$)} in all the line profiles with principal quantum number $n\leq 21$. It is remarkable that these new features abruptly appear when the principal quantum number changes from 22 to 21. This is proven by the increase of 55\% and 80\% of the integrated line fluxes of these two components (see \mbox{Appendix \ref{apendice}).} The increase remains for all the observed RL profiles down to the H15$\alpha$. This is observed in the Hn$\alpha$ RL profiles with $18 \leq n\leq23$ shown in Fig. \ref{H22alpha_and_H21alpha}.

\subsection{Time-variable kinematic features?}
\label{ejection}

The intensity of the two new discovered components probably suffers from time variability because they are affected by laser amplification. However, time variability can be discarded as an explanation for the abrupt appearance of the two new blueshifted kinematic features when the principal quantum number changes from n=22 to n=21. This is supported by the fact that the H22$\alpha$ (without blueshifted components) and the H19$\alpha$, H18$\alpha$, and H15$\alpha$  (with blueshifted components) were observed on the same date (see \mbox{Table \ref{observations}).} Furthermore, the H21$\alpha$ line (with blueshifted components), observed five months after the H22$\alpha$ line, also shows the two blueshifted components, clearly indicating that they are are not due to time variability. In the H21$\alpha$ line profile reported by \citet{Thum1994b}, one can also barely distinguish the blueshifted component of the blue laser peak, but the signal-to-noise ratio was not high enough to confirm the blueshifted component of the red peak. 

\subsection{Origin of the new blueshifted components}

Now we explore the alternative explanation for the drastic kinematic changes in the innermost zones. Since RLs sample deeper zones as $n$ decreases, the two new velocity components indicate that there is a sharp kinematic change in the innermost regions of the disk sampled by the H21$\alpha$ and RLs with lower principal quantum numbers. It is remarkable that the two new velocity components show the same velocity separation, \mbox{$\sim$20 km s$^{-1}$,} relative to the laser peaks, independently of the principal quantum number. This indicates that they might be closely related to the region of the ionized disk from which the laser peaks arise. The most intriguing fact is that both components are blueshifted with respect to their respective laser peaks. The absence of redshifted counterparts might be due to continuum opacity effects that hide the receding material behind the optically thick free-free continuum emission.

\subsection{Ejection of the ionized wind from the disk}

\begin{figure}
\centering
\includegraphics[width=4.8cm]{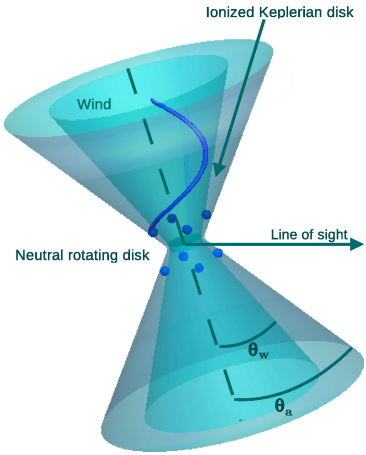}
\caption{Sketch of the geometry assumed in the model, not to scale. The ionized wind was modelled as a double-cone with a semi-opening angle of $\theta_\mathrm{w}=50.5^{\degr}$ against which the ionized disk is placed in the intersected region between the disk wind and the boundary of the neutral disk, given by a double cone with a semi-opening angle of $\theta_\mathrm{a}=57^{\degr}$ \citep{BaezRubio2013}. The electron density distribution assumed in the model depends on the radius as $N_\mathrm{e}\propto r^{-2.14}$. Within the ionized disk, at a radius of $r_\mathrm{l}$, we placed eight symmetric regions depicted as spheres where the disk material is being ejected along the magnetic field lines. The conical helix depicts the trajectory followed by the material ejected from the disk from one of the regions.}
\label{geometry}
\end{figure}

We modelled the observed RL laser emission in the far-IR using the code MORELI \citep{BaezRubio2013} to check the idea that the two new observed blueshifted components might be due to the ionized material ejected from the disk. Figure \ref{geometry} shows a sketch of the assumed geometry for the ionized disk and wind (see also Fig. 3 in \citealt{BaezRubio2013}). Both the ionized wind and disk are assumed to be rotating with a Keplerian law around a massive star of 38 $M_\odot$. The wind has an additional kinematic component of radial expansion at a velocity of \mbox{60 km s$^{-1}$} \citep{BaezRubio2013}. To account for the new velocity components, we included eight clumps symmetrically located within the rotating ionized disk that are assumed to be ejected radially at a velocity of \mbox{35 km s$^{-1}$.} These clumps, approximated by cubes of side \mbox{10 AU}, are located at a distance $r_\mathrm{l}$ from the star in the ionized disk layer, representing the wind-launching radius.

\begin{figure}
\centering
\includegraphics[width=7.6cm]{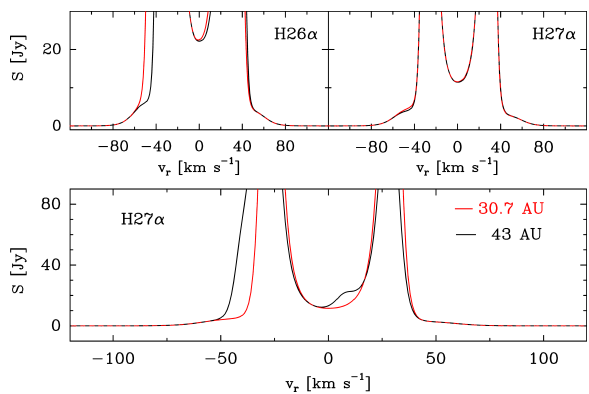}
\caption{Upper panels: comparison of the predicted RL profiles (H26$\alpha$ and H27$\alpha$) by modelling the ionized disk and wind of MWC 349A with and without the ejected clumps (red and black lines) with a velocity of \mbox{35 km s$^{-1}$} in clumps located at a radius of $r_\mathrm{l}=30.7$ AU. 
Lower panel: plot of the predicted H27$\alpha$ profiles assuming that the region from which the wind is launched is located at $r_\mathrm{l}=30.7$ and \mbox{43 AU.}
}
\label{H26alpha_H27alpha_variando_distancia_v4}
\end{figure}

Fitting the strong laser components is beyond the scope of this Letter since MORELI overestimates the velocity-integrated line intensity of RLs because saturation effects are neglected in our model. Since saturation effects are not significant for $n \geq 26$ \citep{BaezRubio2013}, we have considered the H26$\alpha$ and the H27$\alpha$ lines to illustrate the abrupt appearance of kinematic components when the principal quantum number changes by only one unit. The results of the modelling for the parameters derived by \cite{BaezRubio2013} are summarized in Fig. \ref{H26alpha_H27alpha_variando_distancia_v4}. 

We also studied how the predicted RL profiles depend on the radius $r_\mathrm{l}$ where the ejection takes place. The upper panels of \mbox{Fig. \ref{H26alpha_H27alpha_variando_distancia_v4}} show the comparison of the model prediction for the H26$\alpha$ and H27$\alpha$ line profiles for ejected clumps located at \mbox{30.7 AU} (red lines) with the corresponding predicted profiles without clump ejection from the disk (black lines). For this geometry, our model predicts that the two velocity components blueshifted with respect to the laser peak lines appear for the H26$\alpha$ profile (red line of the upper left panel), but not for the H27$\alpha$ profile (red line of the upper right panel). We have found that the radius at which the clumps are located is critical for matching the appearance of the blueshifted components in the H26$\alpha$, but not in the H27$\alpha$. This critical radius is found to be at a continuum optical depth of $\tau_\mathrm{c}\sim3$ for the H26$\alpha$. As expected, the H26$\alpha$ RL emission traces the ejected material unlike the H27$\alpha$ line due to opacity effects. 

The lower panel of \mbox{Fig. \ref{H26alpha_H27alpha_variando_distancia_v4}} shows the predicted line profiles for the H27$\alpha$ when the ejected clumps are located at different radii: $r_\mathrm{l}=$ 30.7 and \mbox{43 AU} (red and black lines). The figure clearly shows that the two blueshifted components only appear for H27$\alpha$ when the ejected material from the disk is located at 43 AU, as expected since in this case the ejected material would be located in the region where the H27$\alpha$ line is optically thin, $\tau_{\mathrm{c}}<3$. It is remarkable that with the present geometry we do not see the redshifted components because of the continuum optical depth. Other geometries for the region from which the ionized gas is ejected, such as an ejection within a ring-like geometry, were explored. However, only the described model with the ejected gas occurring within several clumps is able to reproduce the abrupt appearance of blueshifted components without redshifted counterparts.

We applied this finding to estimate the radius at which the ejection occurs. Our modelling clearly shows that the new detected blueshifted components can abruptly appear when RLs change their principal quantum number by one as long as the ejected material is located in the region where the transition between optically thin and optically thick continuum emission occurs at the frequency of the H21$\alpha$ and H22$\alpha$ lines. From the electron density distribution derived from mm and submm continuum observations \citep{BaezRubio2013} we derived by using MORELI that the optical depth of the continuum emission at the frequency of the H22$\alpha$ line, \mbox{577.9 GHz,} is $\tau_\mathrm{c}=3$ at \mbox{24 AU.} Thus our model predictions qualitatively agree with the observations if the ejected clumps are located at an radius of \mbox{$\sim$24 AU} from the star. This result is consistent with the upper and lower limits provided from the H30$\alpha$ centroid map and the spectral energy distribution analysis \citep{BaezRubio2013}. 

\section{Origin of the wind}

We show clear spectroscopic evidence of the launching of the ionized wind from the disk at a radius smaller than \mbox{24 AU} at velocities of \mbox{$\sim$35 km s$^{-1}$.} This result provides, for the first time, hints about the launching processes of the ionized wind from a disk around a massive star. 

There are two main magnetocentrifugal models predicting wide-angle outflows such as that observed \mbox{toward MWC 349A:} the X-wind models (i.e. \citealt{Shu2000}) and the disk wind models with a steep magnetic field gradient \citep{Pudritz2006}. In particular, in cold disks with winds launched by magnetocentrifugal acceleration, the terminal velocity of the wind along each streamline is $\varv_0\sim\sqrt{2}\varv_\mathrm{l}r_\mathrm{A}/r_\mathrm{l}$ \citep{Konigl2000}, where $\varv_\mathrm{l}=\sqrt{G M}/r_\mathrm{l}$ is the Keplerian velocity at the point on the disk where the wind is launched, and $r_\mathrm{A}$ is the Alfvén radius\footnote{The radius where the density of magnetic energy equals the kinetic energy and beyond which the inertia of the mass loaded in the streamlines prevails over the magnetic interaction.}. The value of $\left(r_\mathrm{A}/r_\mathrm{l}\right)\equiv \sqrt{\lambda}$ is the parameter known as magnetic lever arm because it measures the length along which the wind is magnetocentrifugally accelerated by the poloidal component of the magnetic field \citep{Konigl2000}. Since \mbox{$\varv_0\approx60$ km s$^{-1}$} \citep{BaezRubio2013}, we derive a value of $\sqrt{\lambda}\approx1.25$ using the radial velocity of the ejected material derived with the HIFI RL profiles, \mbox{$\varv_\mathrm{l}=$35 km s$^{-1}$,} close to the sound speed for ionized gas with an electron temperature of \mbox{12000 K} as assumed for MWC 349A \citep{BaezRubio2013}. The derived value of $\lambda$, close to unity, indicates that the wind acceleration occurs along short paths, consistent with our modelling of the ionized gas with two discontinuous kinematic components, a rotating disk, and a wind \citep{BaezRubio2013}. Therefore the acceleration of the wind occurs very close to the region where it is ejected from the disk, and we conclude that the new kinematic component of the wind arising at a radius of \mbox{$\sim$24 AU} is a measure of the radius at which the wind is being ejected from the disk. We note that the value derived for the magnetic lever arm is much lower than that derived from warm disk wind models of T Tauri stars, $\lambda\sim13$ \citep{Casse2000}. This is a logical consequence of the strong mass-loss rate found toward \mbox{MWC 349A,} \mbox{$\sim5\times10^{-5}$ M$_\odot$ yr$^{-1}$,} which is several orders of magnitude higher than those measured toward T Tauri stars, which are typically between $10^{-10}$ and \mbox{$10^{-8}$ M$_\odot$ yr$^{-1}$} \citep{Gullbring1998}.

The question also remains whether X-winds or disk-winds are consistent with the measured radius at which the wind is ejected from the disk. If we assume that the launch is explained by X-winds, the magnetosphere would need to reach radii not much smaller than \mbox{24 AU} to explain our observations. The innermost region of the disk is almost fully ionized because of the high flux of ionizing radiation emitted by the central star. Thus, even if the ionized disk has some magnetic diffusivity, it is an almost perfect conductor and, therefore, the magnetic field lines from the stellar magnetosphere cannot intersect the circumstellar disk at distances from the star much larger than that at which the ionized disk is truncated, that is, at $\sim$0.05 AU \citep{BaezRubio2013}. This finding indicates that an X-wind cannot explain the ejected wind  at the radius found from our results.  In addition, the expected size for the stellar magnetosphere of a massive star can be roughly obtained from the radius at which the magnetic torque on the disc is balanced by the local viscous torque in the disk, as described in \cite{Clarke1995}. Assuming a magnetic field of $\sim$20 kG (one of the strongest magnetic fields measured toward a massive star, \citealt{Wade2012}) in the stellar photosphere (with a radius of $\sim$10 R$_\odot$), and the estimated mass-loss rate and stellar mass for \mbox{MWC 349A,} we derive an upper limit for the size of the magnetosphere of the order of \mbox{1 AU}. This value shows that the stellar magnetosphere, if present, it is much smaller than the required size to explain the observed ionized ejected gas at \mbox{$\sim$24 AU.} We therefore alternatively suggest that an extended magnetic field explains the ejection of the wind as proposed by disk wind models. This hypothesis is also strongly supported by the finding of a dynamically relevant magnetic field of \mbox{$\sim$22 mG} at \mbox{29 AU} \citep{Thum1999,Thum2012}.

\section{Conclusions}

We reported the detection toward MWC 349A of two new blueshifted features by \mbox{$\sim$20 km s$^{-1}$} with respect to the laser peaks in the profiles of Hn$\alpha$ RLs with principal quantum number $15 \leq n\leq 21$. Remarkably, these two features are observed in the H21$\alpha$ line, but not in the RLs with $n\geq22$. Our model show that the abrupt appearance of these two features when the principal quantum number changes from 22 to 21 is consistent with the ionized wind being ejected from the disk in clumps located at radii smaller than \mbox{24 AU.} Our results indicate that magnetohydrodynamic ejection explains the observed ionized wind.

\begin{acknowledgements} 

A. Báez-Rubio thanks CSIC for financial support (JAE predoc 2009 grant). This work has been partially funded by MINECO grants AYA2010-21697-C05-01, AYA2012-32295, \mbox{FIS2012-39162-C06-01,} \mbox{FIS2012-32096,} ESP2013-47809-C3-1-R. Finally, we are also grateful to the anonymous referee and Malcom Wamsley for their valuable comments.
\end{acknowledgements}

\bibliographystyle{aa} 
\bibliography{references}

\begin{appendix}
\section{Quantitative analysis of the sharp appearance of the blueshifted components}
\label{apendice}

The Gaussian decomposition into components is extremely uncertain because of the strong laser peaks. For this reason, we used velocity ranges to describe the different velocity components observed in the RL profiles. For the laser peaks, we use velocity ranges with a width of \mbox{14 km s$^{-1}$,} which are consistent with the width of the Gaussian fit of the laser peaks.  

To quantitatively show the abrupt appearance of the two new revealed blueshifted components with respect to the laser peaks, we compare in this appendix the integrated line fluxes from these two new components and from the two laser peaks (see \mbox{Table \ref{integrated_line_fluxes}).} The integrated line fluxes of the blue and red laser peaks, $S_{\mathrm{(-24,-10)}}$ and $S_{\mathrm{(26,40)}}$, are measured in the radial velocity range between -24 and \mbox{-10 km s$^{-1}$} and 26 and \mbox{40 km s$^{-1}$} . On the other hand, based on visual inspection of the RL profiles \mbox{(Figs. \ref{figure:RLs}} and \ref{H22alpha_and_H21alpha}), we decided to measure the integrated line fluxes of the two new blueshifted components with respect to the laser peaks, $S_{\mathrm{(-50,-30)}}$ and $S_{\mathrm{(10,25)}}$, in radial velocity ranges between -50 and \mbox{-30 km s$^{-1}$} and 10 and \mbox{25 km s$^{-1}$} respectively. 

\begin{figure}[ht!]
\centering
\includegraphics[width=9.25cm]{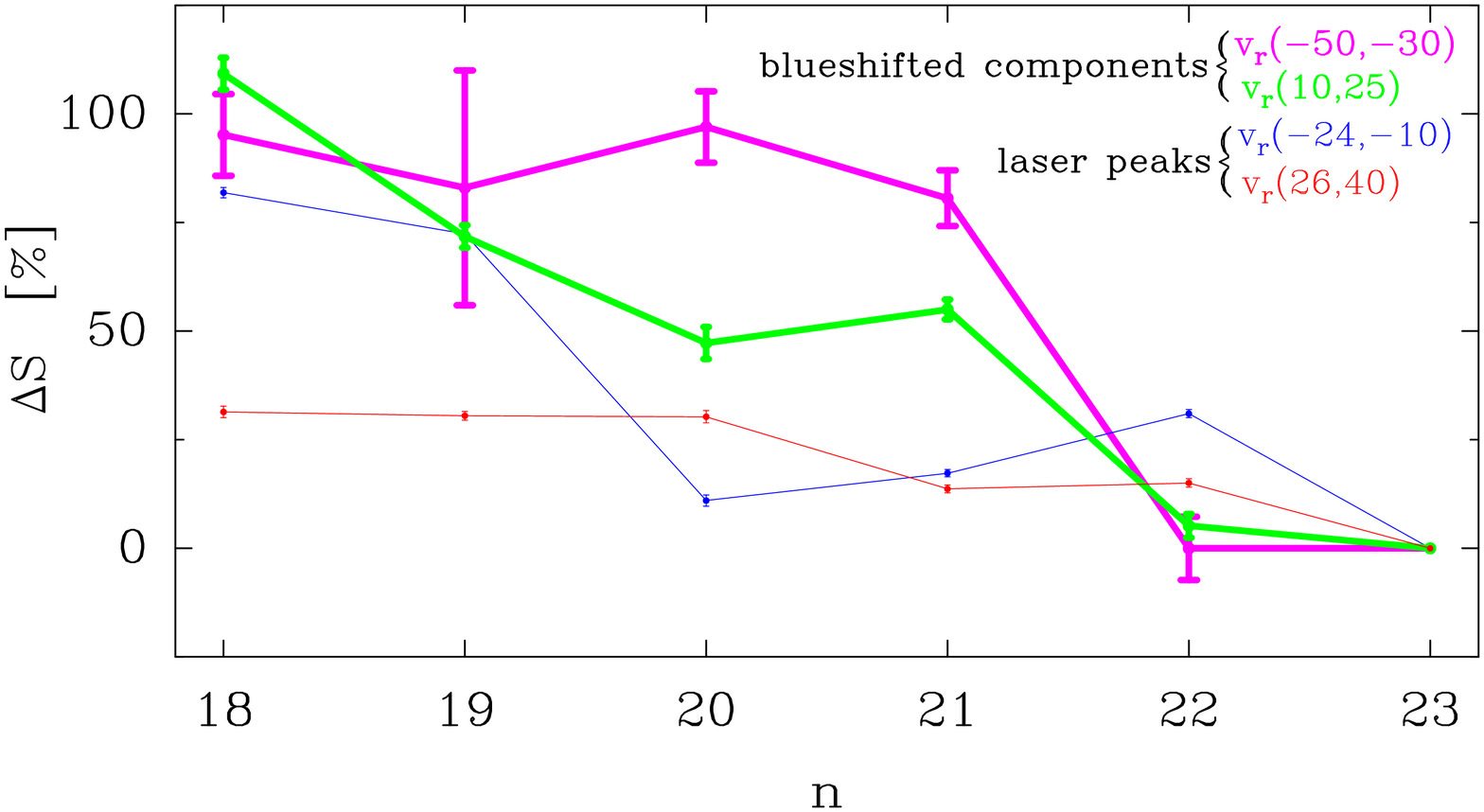}
\caption{Relative percentage increase of the integrated line fluxes of Hn$\alpha$ RLs compared to the H23$\alpha$, $\Delta S$, for four components: the blue and red main laser peaks (blue and red thin lines), and the two new blueshifted components with respect to the blue and red laser peaks (magenta and green thick lines). In the upper right corner we show the radial velocity ranges (in units of \mbox{km s$^{-1}$)} integrated within each component.}
\label{integrated_line_fluxes_increase}
\end{figure}

In \mbox{Fig. \ref{integrated_line_fluxes_increase}} we illustrate the relative increase of the integrated line fluxes of every component with respect to the measured value for the H23$\alpha$, $\Delta S$, as a function of the principal quantum number.  We focus mainly on Hn$\alpha$ RLs with $n\geq18$ because their signal-to-noise ratios for all velocity ranges are higher than one order of magnitude.  This figure clearly shows the trend of the integrated line fluxes of the two new blueshifted components, which are relatively uniform for the H23$\alpha$ and H22$\alpha$ RLs (with a variation smaller than 5.1\%) . This ratio increases to 55\% and 80\% for these velocity components for the H21$\alpha$. Then, the ratios are relatively constant from the H21$\alpha$ to the H18$\alpha$. This behaviour is very different from that shown by the two laser peaks. Unlike the two new blueshifted components, the blue and red laser peaks do not show similar trend as expected, since their intensities are highly sensitive to local physical conditions \citep{BaezRubio2013}. While the integrated line fluxes of the red laser peak increase slowly for decreasing $n$, the blue laser peak does not show any regular trend. Thus we conclude that the sharp increase of the integrated line fluxes of the two new blueshifted components is a clear effect independent of the behaviour of the laser peaks. 

\begin{table}[ht!]
\caption{Measured integrated line fluxes in every velocity range of Hn$\alpha$ RLs with $15\leq n\leq23$.}
\label{integrated_line_fluxes}
\centering
\resizebox{0.5\textwidth}{!}{
\begin{tabular}{ccccc}  \hline
\vspace{-9.5pt}\\
Line& \multicolumn{2}{c}{New blueshifted components}& \multicolumn{2}{c}{Main laser peaks} \\
 & $S_{\mathrm{(-50,-30)}}$ & $S_{\mathrm{(10,25)}}$ & $S_{\mathrm{(-24,-10)}}$ & $S_{\mathrm{(26,40)}}$ \vspace{1pt} \\ 
 & [Jy km s$^{-1}$] & [Jy km s$^{-1}$] &  [Jy km s$^{-1}$]  &  [Jy km s$^{-1}$] \vspace{1pt} \\ \hline \hline 
\vspace{-7.5pt}\\
H23$\alpha$& $330\pm7.9$ & $774\pm6.3$ & $2099.6\pm6.3$ & $2055.5\pm6.3$ \\ 
H22$\alpha$& $330\pm24$ & $814\pm21$ & $2751\pm20$ & $2365\pm20$ \\ 
H21$\alpha$& $596\pm21$ & $1200\pm18$ & $2463\pm18$ & $2337\pm18$ \\  
H20$\alpha$& $560\pm27$ & $1140\pm29$ & $2330\pm28$ & $2678\pm28$ \\
H19$\alpha$& $607\pm27$ & $1330\pm20$ & $3620\pm20$ & $2683\pm20$ \\
H18$\alpha$& $644\pm31$ & $1620\pm29$ & $3820\pm26$ & $2701\pm26$ \\ 
H17$\alpha$& $900\pm96$& $2160\pm79$ & $2714\pm75$ & $2595\pm70$ \\ 
H16$\alpha$& $420\pm79$& $1300\pm63$ & $1780\pm75$ & $1631\pm75$ \\ 
H15$\alpha$& $860\pm120$ & $1940\pm120$ & $3420\pm130$ & $2030\pm130$ \\ \hline 
\end{tabular}
}
\end{table}

\end{appendix}

\end{document}